\documentstyle[preprint,aps]{revtex}

\begin{document}
\title{{\bf $B_{c}$ and heavy meson spectroscopy in the local approximation of the
Schr\"{o}dinger equation with relativistic kinematics}}
\author{Sameer M. Ikhdair\thanks{%
sameer@neu.edu.tr} and \ Ramazan Sever\thanks{%
sever@metu.edu.tr}}
\address{$^{\ast }$Department of Electrical and Electronics Engineering, Near East\\
University, Nicosia, North Cyprus, Mersin 10, Turkey.\\
$^{\dagger }$Department of Physics, Middle East Technical University,\\
Ankara, Turkey.
\\
PACS NUMBER(S): 03.65.Ge, 12.39.Jh, 13.30.Gd}
\date{\today
}
\maketitle

\begin{abstract}
We present bound state masses of the self-conjugate and non-self-conjugate
mesons in the context of the Schr\"{o}dinger equation taking into account
the relativistic kinematics and the quark spins. We apply the usual
interaction by adding the spin dependent correction. The hyperfine
splittings for the ${\rm 2S}$ charmonium and ${\rm 1S}$ bottomonium are
calculated. The pseudoscalar and vector decay constants of the $B_{c}$ meson
and the unperturbed radial wave function at the origin are also calculated.
We have obtained a local equation with a complete relativistic corrections
to a class of three attractive static interaction potentials of the general
form $V(r)=-Ar^{-\beta }+\kappa r^{\beta }+V_{0},$ with $\beta =1,1/2,~3/4$
which can also be decomposed into scalar and vector parts in the form $%
V_{V}(r)=-Ar^{-\beta }+(1-\epsilon )\kappa r^{\beta }$ and $%
V_{S}(r)=\epsilon \kappa r^{\beta }+V_{0};$ where $0\leq \epsilon \leq 1.$
The energy eigenvalues are carried out up to the third order approximation
using the shifted large-N-expansion technique.
\end{abstract}

\begin{center}
$\bigskip $
\end{center}

\begin{verbatim}
 
\end{verbatim}

\section{INTRODUCTION\noindent}

Theoretical interest has risen in the study of the spectroscopy of B$_{c}$
meson in the framework of heavy quarkonium theory [1]. Moreover, the
discovery of the $B_{c}$ \ (the lowest pseudoscalar $^{1}S_{_{0}}$ state)
was reported in 1998 by the Collider Detector at Fermilab (CDF)\
collaboration in $1.8~TeV\ p$-$\overline{p}$\ collisions at the Fermilab [2]
with an observed mass $M_{B_{c}}=6.40\pm 0.39\pm 0.13$ $GeV~$has inspired
new theoretical interest in the subject [3-6]. Further, Kwong and Rosner $%
\left[ 7\right] $ predicted the masses of the lowest vector (triplet) and
pseudoscalar (singlet) states of the $B_{c}$ systems using an empirical mass
formula and a logarithmic potential. Eichten and Quigg $\left[ 1\right] $
calculated the energies and decays of the $B_{c}$ system that was based on
the QCD-motivated potential of Buchm\"{u}ller and Tye $\left[ 8\right] .$
Gershtein {\it et al.} $\left[ 9\right] $ also presented a detailed account
of the energies and decays of the $B_{c}$ system and used a QCD sum-rule
calculations. Baldicchi and Prosperi [6] have computed the $c\overline{b}$
and entire light-heavy quarkonium spectrum based on an effective mass
operator with full relativistic kinematics$.$ Fulcher $\left[ 4\right] $
extended the treatment of the spin-dependent potentials to the full
radiative one-loop level and thus included effects of the running coupling
constant in these potentials. He also used the renormalization scheme
developed by Gupta and Radford $\left[ 10\right] .$ On the other hand,
Motyka and Zalewiski [11] proposed a nonrelativistic potential model to
reproduce the masses of the known $b\overline{b}$ spectrum within the
experimental errors using a new proposed potential form for quarkonia. They
also extended their work [11] by suplementing the Hamiltonian with the
standard spin-dependent terms and produced the $c\overline{c}$ and $c%
\overline{b}$ quarkonium mass spectra, leptonic decay constant and also
decay widths. The shifted large-N expansion technique (SLNET) [12] was
applied to get the spin-averaged data (SAD) of both $Q\overline{Q}$ and $q%
\overline{Q}$ mesons using a group of quarkonium potentials [13] and very
recently was utilized to study the $c\overline{b}~$system in the context of
Schr\"{o}dinger equation and also semi-relativistic quark model [14].

Recently, in 2002, the ALEPH collaboration has searched for the pseudoscalar
bottomonium meson, the $\eta _{b}$ in two-photon interactions at LEP2 with
an integrated luminosity of $699$ pb$^{-1}$ collected at $e^{+}e^{-}$
centre-of mass energies from $181$ ${\rm GeV}$ to 209 ${\rm GeV.}$ One
candidate event is found in the six-charged-particle final state and none in
the four-charged-particle final state. The candidate $\eta _{b}$ ($\eta
_{b}\rightarrow K_{S}K^{-}\pi ^{+}\pi ^{-}\pi ^{+}$) has reconstructed
invariant mass of $9.30\pm 0.02\pm 0.02$ ${\rm GeV}$ [15]${\rm .}$
Theoretical estimates (from perturbative QCD and lattice nonrelativistic QCD
of the mass splitting between $\eta _{b}({\rm 1S)}$ and $\Upsilon ({\rm 1S}%
), $ $M$($\Upsilon ({\rm 1}^{3}{\rm S}_{1}))=9.460$ ${\rm GeV,}$ are
reported (cf. [15] and references therein).

Further, in 2002, the Belle Collaboration [16] has observed a new
pseudoscalar charmonium state, the $\eta _{c}({\rm 2S}),$ in exclusive $%
B\longrightarrow KK_{S}K^{-}\pi ^{+}$ decays. The measured mass of the $\eta
_{c}({\rm 2S}),$ $M$($\eta _{c}({\rm 2S}))=3654\pm 14$ ${\rm MeV.}$ It is
close to the $\eta _{c}({\rm 2S})$ mass observed by the same group in the
experiment $e^{+}e^{-}\longrightarrow J/\psi \eta _{c}$ where $M$($\eta _{c}(%
{\rm 2S}))=3622\pm 12$ ${\rm MeV}$ was found [17]. It is giving rise to a
small hyperfine splitting for the ${\rm 2S}$ state, $\Delta _{{\rm hfs}}(%
{\rm 2S,}$exp)$=M({\rm 2}^{3}{\rm S}_{1})-M({\rm 2}^{1}{\rm S}_{0})=32\pm 14$
${\rm MeV}$ [18]${\rm .}$ Badalian and Bakker [19] calculated the hyperfine
splitting for the ${\rm 2S}$ charmonium state, $\Delta _{{\rm hfs}}({\rm 2S}$%
)$=57\pm 8$ ${\rm MeV,}$ in a recent work. Recksiegel and Sumino developed a
new formalism [20] based on perturbative QCD to compute the hyperfine
splittings of the bottomonium spectrum as well as the fine and hyperfine
splittings of the charmonium spectrum [21].

The motivation of the present calculations is to extend the SLNET [12-14] to
the treatment of the Schr\"{o}dinger equation [13,14] by considering the
spin dependent term $V_{SD}(r)$ that gives the splitting of the singlet and
triplet states and of each $L\geq 1$ level into the four states $^{1}L_{1},$ 
$^{3}L_{L-1},$ $^{3}L_{L}$ and $^{3}L_{L+1}.$ We also present solution for
the Schr\"{o}dinger equation to determine the bound state masses of the $c%
\overline{c},~b\overline{b},~$and $c\overline{b}$ mesons taking into account
the spin-spin, spin-orbit and tensor interactions [22-29]$.$ The spin
effects are treated as perturbation to the static potential. We also
calculate the masses of the recently found new charmonium $\eta _{c}({\rm 2S}%
)$ and the searched bottomonium $\eta _{b}({\rm 1S})$ mesons together with
the hyperfine splittings of their states.

The outline of this paper is as following: In Section II, we first review
briefly the analytic solution of the Schr\"{o}dinger equation for unequal
mass case $(m_{q}\neq m_{Q})$ [14]. Section III is devoted for the class of
three static potentials, which are decomposed into scalar and vector parts
and also for their spin corrections. The cases of pure vector, pure scalar
and equal mixture of vector-scalar coupling interactions are investigated.
The pseudoscalar and vector decay constants of the $B_{c}$ meson are briefly
presented in Section IV. Finally, Section V contains our conclusions.
Appendix A, and B contain some definitions as well as the formulas necessary
to carry out the above mentioned computations.\ \ \ \ \ \ \ \ \ \ \ \ \ \ \
\ \ \ \ \ \ \ \ \ \ \ \ \ \ \ \ \ \ \ \ \ \ \ \ \ \ 

\section{WAVE EQUATION}

We shall consider bound states consisting of fermions with masses $m_{q}$ and%
$~m_{Q}$ and their spins ${\bf S}_{1},~{\bf S}_{2},$ interacting via a
spherically symmetric central potential $V(r).$ Radial part of the
Schr\"{o}dinger equation in the N-dimensional space (in units $\hbar =1)$
[12-14] is:

\begin{equation}
\left\{ -\frac{1}{4\mu }\frac{d^{2}}{dr^{2}}+\frac{[\overline{k}-(1-a)][%
\overline{k}-(3-a)]}{16\mu r^{2}}+V_{eff}(r)\right\} u(r)=E_{n,\ell }u(r),
\label{1}
\end{equation}
where $\mu =\left( m_{q}m_{Q}\right) /(m_{q}+m_{Q})$ denotes the reduced
mass for the two bound interacting particles. Here $E_{n,\ell }$ denotes the
Schr\"{o}dinger binding energy, and $\overline{k}=N+2l-a,$ with $a$
representing a proper shift to be calculated later on and $l$ is the angular
quantum number. We follow the shifted $1/\overline{k}$ expansion method
[13,14] by defining

\begin{equation}
V(r(x))=\frac{\overline{k}^{2}}{Q}\left[ V(r_{0})+\frac{V^{\prime
}(r_{0})r_{0}x}{\bar{k}^{1/2}}+\frac{V^{\prime \prime }(r_{0})r_{0}^{2}x^{2}%
}{2\bar{k}\;}+\cdots \right] ,  \label{2}
\end{equation}
and also the energy eigenvalue expansion [13]

\begin{equation}
E_{n,\ell }=\frac{\overline{k}^{2}}{Q}\left[ E_{0}+E_{1}/\overline{k}+E_{2}/%
\overline{k}^{2}+E_{3}/\overline{k}^{3}+O\left( 1/\overline{k}^{4}\right) %
\right] ,  \label{3}
\end{equation}
where $x=\overline{k}^{1/2}(r/r_{0}-1)$ with $r_{0}$ is an arbitrary point
where the Taylor expansions is being performed about and $Q$ is a scale to
be set equal to $\overline{k}^{2}$ at the end of our calculations. Following
our previous works [13,14], we rewrite down the results as

\begin{equation}
E_{0}=V(r_{0})+\frac{Q}{16\mu r_{0}^{2}},  \label{4}
\end{equation}
\begin{equation}
E_{1}=\frac{Q}{r_{0}^{2}}\left[ \left( n_{r}+\frac{1}{2}\right) \omega -%
\frac{(2-a)}{8\mu }\right] ,  \label{5}
\end{equation}
\begin{equation}
E_{2}=\frac{Q}{r_{0}^{2}}\left[ \frac{(1-a)(3-a)}{16\mu }+\alpha ^{(1)}%
\right] ,  \label{6}
\end{equation}
and 
\begin{equation}
E_{3}=\frac{Q}{r_{0}^{2}}\alpha ^{(2)},  \label{7}
\end{equation}
where $\alpha ^{(1)}$ and $\alpha ^{(2)}$ are listed in Appendix A. Here the
quantity $r_{0}$ is chosen to minimize the leading term, $E_{0}$ [13,14]

\begin{equation}
\frac{dE_{0}}{dr_{0}}=0\text{ \ \ \ and \ \ }\frac{d^{2}E_{0}}{dr_{0}^{2}}>0.
\label{8}
\end{equation}
Therefore, $r_{0}$ satisfies the relation

\begin{equation}
Q=8\mu r_{0}^{3}V^{\prime }(r_{0}),  \label{9}
\end{equation}
and to solve for the shifting parameter $a$, the next contribution to the
energy eigenvalue $E_{1}$ is chosen to vanish [12].

\begin{equation}
a=2-4(2n_{r}+1)\mu \omega ,  \label{10}
\end{equation}
with

\begin{equation}
\omega =\frac{1}{4\mu }\left[ 3+\frac{r_{0}V^{\prime \prime }(r_{0})}{%
V^{\prime }(r_{0})}\right] ^{1/2}.  \label{11}
\end{equation}
Once $r_{0}$ is being determined, with the choice $\overline{k}=\sqrt{Q}$
which rescales the potential, we get an analytic expression the energy
eigenvalues (3). The Coulomb potential is considered as a testing case, the
results are found to be strongly convergent and highly accurate. The
calculations of the energy eigenvalues were carried out up to the second
order correction. Therefore, the bound state energy to the third order
becomes

\begin{equation}
E_{n,l}=E_{0}+\frac{1}{r_{0}^{2}}\left[ \frac{(1-a)(3-a)}{16\mu }+\alpha
^{(1)}+\frac{\alpha ^{(2)}}{\overline{k}}+O\left( \frac{1}{\overline{k}^{2}}%
\right) \right] .  \label{12}
\end{equation}
Once the problem is collapsed to its actual dimension $N=3,$ it simply rests
the task of relating the coefficients of our equation to the one-dimensional
anharmonic oscillator in order to read the energy spectrum. One obtains

\begin{equation}
1+2l+(2n_{r}+1)\left[ 3+\frac{r_{0}V^{\prime \prime }(r_{0})}{V^{\prime
}(r_{0})}\right] ^{1/2}=\left[ 8\mu r_{0}^{3}V^{\prime }(r_{0})\right]
^{1/2}.  \label{13}
\end{equation}
We finally write the bound state mass for spinless particles as

\begin{equation}
M(q\overline{Q})=m_{q}+m_{Q}+2E_{n,l}.  \label{14}
\end{equation}
where $m_{q}$ and $m_{Q}$ are the constituent meson masses whereas $%
n=n_{r}+1 $ is the principal quantum number. As stated before [13,14], for a
fixed $n$ the computed energies become more accurate as $l$ increases. This
is expected since the expansion parameter $1/\overline{k}$ becomes smaller
as $l $ becomes larger since the parameter $\overline{k}$ is proportional to 
$n$ and appears in the denominator in higher-order correction.

\section{HEAVY QUARKONIUM AND $B_{c}$ MESON MASS SPECTRA\noindent}

The spin-independent potential (which may be velocity dependent) essentially
yields SAD. Furthermore, the spin-dependent term $V_{SD}(r)$ gives the
splitting both of the $^{3}S_{1}$ and \ $^{1}S_{0}$, with ${\bf S}={\bf S}%
_{1}+{\bf S}_{2}$ is $1$ and $0$ for triplet and singlet states,
respectively, and of each level into the four states $^{3}L_{L-1},$ $%
^{3}L_{L},$ $^{3}L_{L+1}$ and $^{1}L_{1}.$ Thus the potential takes
[23,26,28-30] 
\begin{equation}
V_{eff}(r)=V_{static}(r)+V_{SD}(r)+V_{SI}(r),  \label{15}
\end{equation}
with spin-dependent and spin-independent perturbation terms are given in
Refs. [26,28,30]. Further, the static potential [14,31] takes the general
form

\begin{equation}
V_{static}(r)=-Ar^{-\beta }+\kappa r^{\beta }+V_{0};~\beta =1,1/2,~3/4,\text{
}A,\kappa \geq 0  \label{16}
\end{equation}
which has a limited character of Ref. [11,32], (i.e., same $\beta ),$ where $%
V_{0}$ may be of either sign. The form (16) includes three types of static
potentials. The first static potential we consider is the Cornell [33]
potential $(\beta =1)$ which is one of the earliest QCD-motivated potentials
in the literature

\begin{equation}
V_{C}(r)=-\frac{A}{r}+\kappa r+V_{0},  \label{17}
\end{equation}
where $\ A=4\alpha _{s}/3,~$is a short range gluon exchange, and $\kappa ~$%
is a confinement constant. The second potential is that of Song and Lin [34] 
$(\beta =1/2)$ which is given by 
\begin{equation}
V_{S-L}(r)=-\frac{A}{r^{1/2}}+\kappa r^{1/2}+V_{0}.  \label{18}
\end{equation}
The third potential is an intermediate case between the last mentioned
potentials and is called Turin potential [31] $(\beta =3/4)$ which has the
form

\begin{equation}
V_{T}(r)=-\frac{A}{r^{3/4}}+\kappa r^{3/4}+V_{0}.  \label{19}
\end{equation}
The class of static potentials in Eq. (16) must satisfy the following
conditions [31]

\begin{equation}
\frac{dV}{dr}>0,~\frac{d^{2}V}{dr^{2}}\leq 0.  \label{20}
\end{equation}
On the other hand, the expression (16) can be rewritten in a more general
form with two different power parameters $\alpha $ and $\beta $ as [32]:

\begin{equation}
V(r)=-Ar^{-\alpha }+\kappa r^{\beta }+V_{0},~  \label{21}
\end{equation}
where $\alpha \neq \beta .$ Motyka and Zalewiski [11] utilized the form (21)
by setting $\alpha =1$ and $\beta =1/2;$ that is,

\begin{equation}
V(r)=-\frac{A}{r}+\kappa \sqrt{r}+V_{0},~  \label{22}
\end{equation}
The potential form (22) belongs to the class of generality (21) and was
successfuly used by Motyka {\it et al.} in fitting the $c\overline{c}$
spectrum and later on extended to the $b\overline{b}$ and $B_{c}$
spectroscopy [11]. In this work we devote our study to the first class of
generality (16) leaving the second class of generality (21) for further
study. We will use a fairly flexible parameterization of the potentials of
(16) in fitting the data and take the nonrelativistic interaction as a sum
of scalar and vector terms as it follows from the Lorentz invariance theory
[26,29]

\begin{equation}
V_{V}(r)=-Ar^{-\beta }+(1-\epsilon )\kappa r^{\beta },  \label{23}
\end{equation}
and

\begin{equation}
V_{S}(r)=\epsilon \kappa r^{\beta }+V_{0},  \label{24}
\end{equation}
where $\epsilon $ is the mixing coefficient. The vector term incorporates
the expected short-distance behavior from single-gluon exchange. We have
also included a multiple of the long-range interaction in $V_{V}(r)$ to see
the nature of the confining interaction. Here, we investigate the cases of
pure scalar confinement $(\epsilon =1),$ equal mixture of scalar-vector
couplings $(\epsilon =1/2)$ and a pure vector case $(\epsilon =0).$

The total spin-dependent potential $V_{SD}(r)$ given by [23-25,28-30]

\[
V_{SD}(r)=V_{A}+V_{S}=\frac{1}{4}\left[ \frac{1}{m_{q}^{2}}-\frac{1}{%
m_{Q}^{2}}\right] \left[ \frac{V_{V}^{\prime }(r)-V_{S}^{\prime }(r)}{r}%
\right] {\bf L}\cdot {\bf S}_{-} 
\]

\[
+\frac{{\bf L}\cdot {\bf S}}{m_{q}m_{Q}}\frac{V_{V}^{\prime }(r)}{r}+\frac{1%
}{2}\left[ \frac{{\bf L}\cdot {\bf S}_{1}}{m_{q}^{2}}+\frac{{\bf L}\cdot 
{\bf S}_{2}}{m_{Q}^{2}}\right] \left[ \frac{V_{V}^{\prime }(r)-V_{S}^{\prime
}(r)}{r}\right] 
\]
\begin{equation}
+\frac{2}{3}\frac{{\bf S}_{1}\cdot {\bf S}_{2}}{m_{q}m_{Q}}\left[ \nabla
^{2}V_{V}(r)\right] +\frac{S_{12}}{m_{q}m_{Q}}\left[ -V_{V}^{\prime \prime
}(r)+\frac{V_{V}^{\prime }(r)}{r}\right] ,  \label{25}
\end{equation}
where ${\bf S}_{1}$ and ${\bf S}_{2}$ are the quark spins, ${\bf S}_{-}={\bf %
S}_{1}-{\bf S}_{2},$ ${\bf L=x\times p}$ is the relative orbital angular
momentum, and $S_{12}=T-\left( {\bf S}_{1}\cdot {\bf S}_{2}\right) /3$ where 
$T=({\bf S}_{1}\cdot \widehat{{\bf r}})({\bf S}_{2}\cdot \widehat{{\bf r}})$
\ is the tensor operator with the versor $\widehat{{\bf r}}={\bf r/}r$ . The
spin dependent correction (25), which is responsible for the hyperfine
splitting of the mass levels, in the short-range is generally used in the
form for $S$-wave $(L=0)$ (cf. e.g., [23,30]): 
\begin{equation}
V_{hfs}(r)=\frac{2}{3}({\bf S}_{1}\cdot {\bf S}_{2})\nabla ^{2}\left[ -\frac{%
4\alpha _{s}}{3r^{\beta }}\right] ,  \label{26}
\end{equation}
and the one responsible for the fine splittings used for $P$- and $D$-waves $%
(L\neq 0)$ is:

\[
V_{fs}(r)=\frac{1}{m_{q}m_{Q}}\left\{ \frac{{\bf L}\cdot {\bf S}}{r}\left[
\left( 1+\frac{1}{4}\frac{m_{q}^{2}+m_{Q}^{2}}{m_{q}m_{Q}}\right)
V_{V}^{\prime }(r)-\frac{1}{4}\frac{m_{q}^{2}+m_{Q}^{2}}{m_{q}m_{Q}}%
V_{S}^{\prime }(r)\right] \right. 
\]

\begin{equation}
+\frac{2}{3}({\bf S}_{1}\cdot {\bf S}_{2})\nabla ^{2}\left[ \kappa
(1-\epsilon )r^{\beta }\right] +\left. \left[ T-\frac{1}{3}\left( {\bf S}%
_{1}\cdot {\bf S}_{2}\right) \right] \left[ V_{V}^{\prime \prime }(r)+\frac{%
V_{V}^{\prime }(r)}{r}\right] \right\} ,  \label{27}
\end{equation}
where the matrix element can be evaluated in terms of the expectation values 
$\left\langle {\bf L}\cdot {\bf S}_{1}\right\rangle =\left\langle {\bf L}%
\cdot {\bf S}_{2}\right\rangle =\frac{1}{2}\left\langle {\bf L}\cdot {\bf S}%
\right\rangle $. Hence, Eq. (27) is the complete spin-dependent potential in
QCD through order $m^{2}.$ For bound state constituents of spin $%
S_{1}=S_{2}=1/2,$ the scalar product of their spins ${\bf S}_{1}\cdot {\bf S}%
_{2}$ and ${\bf L}\cdot {\bf S}$ are to be found in the Appendix B. The
appearance of a Coulomb-like contribution $\sim 1/r$ in the vector part of
the potential causes some problems due to the relation $\nabla
^{2}(1/r)=-4\pi \delta ^{(3)}(x),$ in the spin-spin interaction (26)
involves a delta function of the $S$-wave $(L=0).$ Thus, for Cornell
potential, the hyperfine splitting potential (26) gives 
\begin{equation}
V_{eff}(r)=-\frac{A}{r}+\kappa r+\frac{32\pi \alpha _{s}}{9m_{q}m_{Q}}\delta
^{(3)}({\bf r})({\bf S}_{1}\cdot {\bf S}_{2})+V_{0};\text{ where }\beta =1.
\label{28}
\end{equation}
Therefore for the energy of spin-spin interaction we have approximately:

\begin{equation}
E_{ss}=\frac{1}{2M_{n,0}}\Delta M_{ss}^{2}\left\langle {\bf S}_{1}\cdot {\bf %
S}_{2}\right\rangle ,  \label{29}
\end{equation}
where $M_{n,0}$ is given in Eq. (14) and the singlet-triplet mass squared
difference 
\begin{equation}
\Delta M_{ss}^{2}=M_{S=1}^{2}-M_{S=0}^{2}\simeq \frac{32}{9}\alpha
_{s}\kappa ,  \label{30}
\end{equation}
for light $q\overline{q}$ systems (in the instantaneous-limit approximation)
[23], and 
\begin{equation}
\Delta M_{ss}^{2}=M_{S=1}^{2}-M_{S=0}^{2}\simeq \frac{256}{3\pi ^{2}}\alpha
_{s}\kappa ,  \label{31}
\end{equation}
for heavy quarkonia (hydrogen-like trial functions) [23]. All these
predictions for the mass-squared difference are independent of the mass of
the particles which constitute the bound state. Further, for the Song-Lin
and Turin potentials, it also give

\begin{equation}
V_{eff}(r)=-\frac{A}{r^{\beta }}+\kappa r^{\beta }+\frac{8\beta (1-\beta
)\pi \alpha _{s}}{9m_{q}m_{Q}r^{2}}r^{-\beta }{\bf S}_{1}\cdot {\bf S}%
_{2}+V_{0};\text{ where }\beta =1/2,3/4.  \label{32}
\end{equation}
Like most authors (cf. [1]), we determine the coupling constant $\alpha
_{s}(m_{c}^{2})$ from the well measured hyperfine splitting for the ${\rm 1S}%
(\overline{c}c)$ state [18]

\begin{equation}
\Delta E_{{\rm HF}}({\rm 1S},\exp )=M_{J/\psi }-M_{\eta _{c}}=117.2\pm 1.5~%
{\rm MeV},  \label{33}
\end{equation}
and also for the ${\rm 2S}(\overline{c}c)$ state [16-19] 
\begin{equation}
\Delta E_{{\rm HF}}({\rm 2S},\exp )=M_{\psi ^{\prime }}-M_{\eta _{c^{\prime
}}}=32\pm 14~{\rm MeV},  \label{34}
\end{equation}
for each desired potential. On the other hand, the Eq. (27), for $P$, $%
D,\cdots $ waves $(L\neq 0)$ case, gives

\begin{equation}
V_{eff}(r)=V_{static}(r)+g(r)\left[ F_{LS_{-}}\left( {\bf L}\cdot {\bf S}%
_{-}\right) +F_{LS}\left( {\bf L}\cdot {\bf S}\right) +F_{SS}\left( {\bf S}%
_{1}\cdot {\bf S}_{2}\right) +F_{T}T\right] ,  \label{35}
\end{equation}
with a given set of spin-dependent quantities 
\begin{equation}
F_{LS_{-}}=\left[ \frac{1}{4}\frac{m_{Q}^{2}-m_{q}^{2}}{m_{q}m_{Q}}\left[
Ar^{-\beta }+\left( 1-\epsilon \right) \kappa r^{\beta }\right] \right] ,
\label{36}
\end{equation}

\begin{equation}
F_{LS}=\left[ \left( 1+\frac{1}{4}\frac{m_{q}^{2}+m_{Q}^{2}}{m_{q}m_{Q}}%
\right) \left[ Ar^{-\beta }+\left( 1-2\epsilon \right) \kappa r^{\beta }%
\right] +\epsilon \kappa r^{\beta }\right] ,  \label{37}
\end{equation}

\begin{equation}
F_{SS}=\left[ -\frac{(2+\beta )}{3}Ar^{-\beta }+\beta \left( 1-\epsilon
\right) \kappa r^{\beta }\right] ,  \label{38}
\end{equation}
and

\begin{equation}
F_{T}=\left[ (2+\beta )Ar^{-\beta }+(2-\beta )\left( 1-\epsilon \right)
\kappa r^{\beta }\right] ,  \label{39}
\end{equation}
where $g(r)=\frac{\beta }{m_{q}m_{Q}r^{2}}$ is a necessary coupling function$%
.$ The spin-independent corrections in Eq. (15) are explicitly given in
Refs. [28,30] which are not treated in our present work.

\subsection{Singlet states}

For parastates $\left( L=J\right) $ or $(S=0)$ case, we have parity $%
P=(-1)^{J+1}$ and charge conjugation $C=(-1)^{L}.$ Thus, the potential (35)
can be rewritten as

\begin{equation}
V_{eff}(r)=V_{static}(r)-\frac{1}{4}\left( 3F_{SS}+F_{T}\right) +\sqrt{\frac{%
1}{10}(2L+3)(2L-1)}F_{LS_{-}},  \label{40}
\end{equation}
which can be substituted in Eq. (1) and also by setting $\overline{k}=N+2J-a$
therein$.$ Further, Eqs. (28) and (32) give 
\begin{equation}
V_{eff}(r)=-\frac{A}{r}+\kappa r-\frac{8\pi \alpha _{s}}{3m_{q}m_{Q}}\delta
^{(3)}({\bf r})+V_{0},  \label{41}
\end{equation}
and 
\begin{equation}
V_{eff}(r)=-\frac{A}{r^{\beta }}+\kappa r^{\beta }-\frac{2\beta (1-\beta
)\pi \alpha _{s}}{3m_{q}m_{Q}r^{2}}r^{-\beta }+V_{0};\text{ where }\beta
=1/2,3/4,  \label{42}
\end{equation}
respectively, which generate singlet states with opposite quark and
antiquark spins of the signature $n^{1}S_{0}.$ Furthermore, Eq. (40) can be
rewritten simply as

\[
V_{J=L}(r)=g(r)\left\{ \frac{1}{4}\frac{m_{Q}^{2}-m_{q}^{2}}{m_{q}m_{Q}}%
\sqrt{\frac{1}{10}(2L+3)(2L-1)}\left[ Ar^{-\beta }+(1-\epsilon )\kappa
r^{\beta }\right] \right. 
\]
\begin{equation}
-\left. \frac{1}{2}(1+\beta )(1-\epsilon )\kappa r^{\beta }\right\}
+V_{static}(r)  \label{43}
\end{equation}
which generates states of the signatures $%
n^{1}P_{1},~n^{1}D_{2},~n^{1}F_{3},~n^{1}G_{4},\cdots .$\ \ \ \ \ \ \ \ \ \
\ \ \ \ \ \ \ \ \ \ \ \ \ \ \ \ \ \ \ \ \ \ \ \ \ \ \ \ \ \ \ \ \ \ \ \ \ \
\ \ \ \ \ \ \ \ \ \ \ \ \ \ \ \ \ \ \ \ \ \ \ \ \ \ \ \ \ \ \ \ \ \ \ \ \ \
\ \ \ \ \ \ \ \ \ \ \ \ \ \ \ \ \ \ \ \ \ \ \ \ \ \ \ \ \ \ \ \ \ \ \ \ \ \
\ \ \ \ \ \ \ \ \ \ \ \ \ \ \ \ \ \ \ \ \ \ \ \ \ \ \ \ \ \ \ \ \ \ \ \ \ \
\ \ \ \ \ \ \ \ \ \ \ \ \ \ \ \ \ \ \ \ \ \ \ \ \ \ \ \ \ \ \ \ \ \ \ \ \ \
\ \ \ \ \ \ \ \ \ \ \ \ \ \ \ \ \ \ \ \ \ \ \ \ \ \ \ \ \ \ \ \ \ \ \ \ \ \
\ \ \ \ \ 

\subsection{Triplet states}

For triplet $(S=1)$ case, we have the known inequality $\left| L-S\right|
\leq J\leq L+S$ that gives $J=L$ and $~L\pm 1:$ \ \ \ \ \ \ \ \ \ \ \ \ \ \
\ \ \ \ \ \ \ \ \ \ \ \ \ \ \ \ \ \ \ \ \ \ \ \ \ \ \ \ \ \ \ \ \ \ \ \ \ \
\ \ \ \ 

\subsubsection{States $J=L$}

Here, the parity $P=(-1)^{J+1}$and the charge conjugation $C=(-1)^{L+1}.$
The potential in Eq. (35) takes the following simple form

\begin{equation}
V_{eff}(r)=V_{static}(r)+\frac{1}{4}\left( F_{SS}+F_{T}-4F_{LS}\right) +%
\sqrt{\frac{1}{10}(2L+3)(2L-1)}F_{LS_{-}},  \label{44}
\end{equation}
which can be substituted in (1) together with $\overline{k}=N+2J-a$ therein$%
. $ Further, the potential (44) reads

\[
V_{J=L}(r)=-\frac{g(r)}{2}\left\{ \left[ \left( \frac{4-\beta }{3}+\frac{1}{2%
}\frac{m_{q}^{2}+m_{Q}^{2}}{m_{q}m_{Q}}\right) Ar^{-\beta }+\left( 1+\frac{1%
}{2}\frac{m_{q}^{2}+m_{Q}^{2}}{m_{q}m_{Q}}\right) (1-2\epsilon )\kappa
r^{\beta }+\epsilon \kappa r^{\beta }\right] \right. 
\]

\begin{equation}
\left. -\frac{1}{2}\frac{m_{Q}^{2}-m_{q}^{2}}{m_{q}m_{Q}}\sqrt{\frac{1}{10}%
(2L+3)(2L-1)}\left[ Ar^{-\beta }+(1-\epsilon )\kappa r^{\beta }\right]
\right\} +V_{static},  \label{45}
\end{equation}
which generates states like $n^{3}P_{1},~n^{3}D_{2},~n^{3}F_{3},~n^{3}G_{4},%
\cdots .$

\subsubsection{States $J=L\pm 1$}

We have the parity $P=(-1)^{J}$ and the charge conjugation $C=(-1)^{L+1}.$
The eigenfunction is a superposition of two components with orbital momentum 
$L=J+1$ and $L=J-1$ which have equal space parity

\begin{equation}
\psi _{S,J}(r)=u_{J-1}(r)Y_{J-1,1,J}^{m}(\theta ,\varphi
)+u_{J+1}(r)Y_{J+1,1,J}^{m}(\theta ,\varphi ).  \label{46}
\end{equation}
The action of the tensor operator, $T,$ on the two components of the
wavefunction in Eq. (46) is 
\begin{equation}
Tu_{J\pm 1}Y_{J\pm 1,1,J}^{m}(\widehat{{\bf r}})=\mp \frac{1}{4(2J+1)}%
u_{J\pm 1}Y_{J\pm 1,1,J}^{m}(\widehat{{\bf r}})+\frac{1}{2}\frac{\sqrt{J(J+1)%
}}{2J+1}u_{J\mp 1}Y_{J\mp 1,1,J}^{m}(\widehat{{\bf r}}).  \label{47}
\end{equation}
Therefore, a set of equations are obtained 
\[
\left\{ -\frac{1}{4\mu }\frac{d^{2}}{dr^{2}}+\frac{\left[ \overline{k}%
-\left( 1-a\right) \right] \left[ \overline{k}-\left( 3-a\right) \right] }{%
16\mu r^{2}}+V_{static}(r)-E_{n,J+1}-\left( J+2\right) F_{LS}\right. 
\]

\begin{equation}
\left. +\frac{1}{4}\left( F_{SS}-\frac{F_{T}}{\left( 2J+1\right) }\right)
\right\} u_{n,J+1}(r)=\frac{\sqrt{J(J+1)}}{2\left( 2J+1\right) }F_{T}\text{ }%
u_{n,J-1}(r),  \label{48}
\end{equation}
and

\[
\left\{ -\frac{1}{4\mu }\frac{d^{2}}{dr^{2}}+\frac{\left[ \overline{k}%
-\left( 1-a\right) \right] \left[ \overline{k}-\left( 3-a\right) \right] }{%
16\mu r^{2}}+V_{static}(r)-E_{n,J-1}+\left( J-1\right) F_{LS}\right. 
\]

\begin{equation}
\left. +\frac{1}{4}\left( F_{SS}+\frac{F_{T}}{\left( 2J+1\right) }\right)
\right\} u_{n,J-1}(r)=\frac{\sqrt{J(J+1)}}{2\left( 2J+1\right) }F_{T}\text{ }%
u_{n,J+1}(r),  \label{49}
\end{equation}
where $\overline{k}=N+2J+2-a.$ Therefore, Eqs. (48) and (49) describe states
such as $%
n^{3}P_{2},~n^{3}D_{3},~n^{3}F_{2},~n^{3}H_{4},n^{3}P_{0},~n^{3}D_{1},\cdots
.$ Here we may consider numerically the system obtained and separate
equations by dropping out the mixed terms to see their effect on the
spectrum of the masses. Consequently one can rewrite (48) and (49) in the
following simplest forms

\[
V_{J=L-1}(r)=-g(r)\left\{ (L+1)\left[ \left( 1+\frac{1}{4}\frac{%
m_{q}^{2}+m_{Q}^{2}}{m_{q}m_{Q}}\right) \left[ Ar^{-\beta }+\left(
1-2\epsilon \right) \kappa r^{\beta }\right] +\epsilon \kappa r^{\beta }%
\right] \right. 
\]

\[
+\frac{1}{4}\frac{1}{(2L-1)}\left[ (2+\beta )Ar^{-\beta }+(2-\beta )\left(
1-\epsilon \right) \kappa r^{\beta }\right] 
\]
\begin{equation}
\left. +\frac{1}{4}\left[ \frac{(2+\beta )}{3}Ar^{-\beta }-\beta \left(
1-\epsilon \right) \kappa r^{\beta }\right] \right\} +V_{static}(r),
\label{50}
\end{equation}
for states $n^{3}P_{0},~n^{3}D_{1},~n^{3}F_{2},~n^{3}H_{4},\cdots $ and

\[
V_{J=L+1}(r)=g(r)\left\{ \left[ \left( 1+\frac{1}{4}\frac{m_{q}^{2}+m_{Q}^{2}%
}{m_{q}m_{Q}}\right) \left[ Ar^{-\beta }+\left( 1-2\epsilon \right) \kappa
r^{\beta }\right] +\epsilon \kappa r^{\beta }\right] L\right. 
\]

\[
+\frac{1}{4}\frac{1}{(2L+3)}\left[ (2+\beta )Ar^{-\beta }+(2-\beta )\left(
1-\epsilon \right) \kappa r^{\beta }\right] 
\]
\begin{equation}
\left. -\frac{1}{4}\left[ \frac{(2+\beta )}{3}Ar^{-\beta }-\beta \left(
1-\epsilon \right) \kappa r^{\beta }\right] \right\} +V_{static}(r),
\label{51}
\end{equation}
for states $n^{3}P_{2},~n^{3}D_{3},\cdots .$ Further, for triplet $S$-wave,
we have 
\begin{equation}
V_{eff}(r)=-\frac{A}{r}+\kappa r+\frac{8\pi \alpha _{s}}{9m_{q}m_{Q}}\delta
^{(3)}({\bf r})+V_{0},  \label{52}
\end{equation}
and 
\begin{equation}
V_{eff}(r)=-\frac{A}{r^{\beta }}+\kappa r^{\beta }+\frac{2\beta (1-\beta
)\pi \alpha _{s}}{9m_{q}m_{Q}r^{2}}r^{-\beta }+V_{0};\text{ where }\beta
=1/2,3/4,  \label{53}
\end{equation}
which describe states such as $n^{3}S_{1.}$

\subsubsection{State $J=0$}

Equations (50) and (51) degenerate into a single equation with an effective
potential

\begin{equation}
V_{eff}(r)=V_{static}(r)+\frac{1}{4}(F_{SS}-F_{T}-8F_{LS}),  \label{54}
\end{equation}
and also by setting $\overline{k}=N+2-a$ therein$.$ Further, Eq. (54) becomes

\[
V_{J=0}(r)=-\left\{ 1+g(r)\left[ 2+\frac{1}{2}\frac{m_{q}^{2}+m_{Q}^{2}}{%
m_{q}m_{Q}}+\frac{2+\beta }{3}\right] \right\} Ar^{-\beta } 
\]

\begin{equation}
+\left\{ 1-\frac{g(r)}{2}\left[ \left( 5-\beta \right) \left( 1-\epsilon
\right) +\left( \frac{m_{q}^{2}+m_{Q}^{2}}{m_{q}m_{Q}}\right) \left(
1-2\epsilon \right) \right] \right\} \kappa r^{\beta }+V_{0}.  \label{55}
\end{equation}
which only describes states such as $n^{3}P_{0}.$

\section{PSEUDOSCALAR AND VECTOR DECAY CONSTANTS OF THE $B_{c}$ MESON}

The $B_{c}$ can decay via electromagnetic and pionic transitions into the
lightest pseudoscalar ground state $B_{c}.$ The significant contribution to
the $B_{c} $ total decay rate comes from the annihilation of the $c$ quark
and $\overline{b}$ antiquark into the vector boson $W^{+}$ which decays into
a lepton and a neutrino or a quark-antiquark pair. The weak annihilation
decay rate is determined by the pseudoscalar constant of the $B_{c}$ meson.

The nonrelativistic expression for the decay constants is given by [35-37]

\begin{equation}
f_{P}^{NR}=f_{V}^{NR}=\sqrt{\frac{12}{M_{P,V}(q\overline{Q})}}\left| \Psi
_{P,V}(0)\right| ,  \label{56}
\end{equation}
where $\Psi _{P,V}(0)$ is the meson wave function at the origin ($r=0),$ $%
f_{P}$ and $f_{V}$, $P$ corresponds to the pseudoscalar $B_{c}$ and $V$ \ to
to the vector $B_{c}^{\ast }$ mesons and $M_{P,V}(q\overline{Q})$ are the
masses of the $B_{c}$ and $B_{c}^{\ast }$ mesons.

\section{RESULTS AND CONCLUSIONS}

We have given further tests for the potential model in the context of
Schr\"{o}dinger equation using SLNET and also extended our earlier formalism
for the SAD spectra [14] to the calculation of all the states by introducing
the spin corrections. The obtained mass formula (14) include fine and
hyperfine splitting of the energy levels. This mass formula is able to
describe with some accuracy the spectra of all quark-antiquark bound states.
We have obtained the self-conjugate meson spectroscopy using a group of
three static potential model. This model has also been extended to comprise
various cases of pure scalar confinement $(\epsilon =1),$ scalar-vector
couplings $(\epsilon =1/2)$ and the vector confinement $(\epsilon =0)$
interactions. The parameters used are shown in Table I. Our results for mass
spectrum of $c\overline{c},$ $b\overline{b},$ and $c\overline{b}$ systems
with the static potentials, in the flavour-independent case are presented in
Tables II-IV. Different sets of parameters for the Cornell potential are
used to produce the binding masses of heavy quarkonium states as shown in
Tables V. Further, the $c\overline{b}$ mass spectrum is given in Table VI.
In the equal scalar and vector couplings, we have found that our fits are
very good with level values and accurate to a few ${\rm MeV}.$ For
convenience we compare explicitly the predicted and measured spin splitting
energy for different $L$ states. We find that the apparent success is
achieved for the predicted $\chi _{b2}-\chi _{b1}=24~{\rm MeV}$ and $\chi
_{b1}-\chi _{b0}=33~{\rm MeV}$ in the average for the three potentials and
are very close to the experimental values $21~{\rm MeV}$ and 32$~{\rm MeV}$
respectively. Furthermore, the predicted $\chi _{b2}^{\prime }-\chi
_{b1}^{\prime }=13~{\rm MeV}$ and $\chi _{b1}^{\prime }-\chi _{b0}^{\prime
}=16~{\rm MeV}$ in the average for the three potentials which are exactly
same as the experimental value $13~{\rm MeV}$ and close to 2$3~{\rm MeV},$
respectively. The predicted hyperfine splitting $\Delta _{{\rm hfs}}{\rm (1S)%
}=M(\Upsilon {\rm (1S)})-M(\eta _{b}{\rm (1S)})=80_{-8}^{+6}~{\rm MeV,}$
(cf. [15])$,$ $\Delta _{{\rm hfs}}{\rm (2S)}=M(\Upsilon ^{\prime }{\rm (2S)}%
)-M(\eta _{b}^{\prime }{\rm (2S)})=22_{-2}^{+3}~{\rm MeV,}$ and $\Delta _{%
{\rm hfs}}{\rm (3S)}=M(\Upsilon ^{\prime \prime }{\rm (3S)})-M(\eta
_{b}^{\prime \prime }{\rm (3S)})=14_{-1}^{+2}~{\rm MeV}$ are nearly close to
the theoretically calculated values 62$~{\rm MeV},$ $40$ ${\rm MeV,}$ and $15
$ ${\rm MeV,}$ respectively, (cf. [15,21]). Further, The hyperfine splitting
for the ${\rm 2S}$ charmonium state is calculated and the predicted number
is $\Delta _{{\rm hfs}}{\rm (2S)}=M(\psi {\rm (2S)})-M(\eta _{c}{\rm (2S)}%
)=56_{-8}^{+4}~{\rm MeV}$ for flavour dependent case and $56_{-8}^{+18}~{\rm %
MeV}$ for flavour independent case, (cf. [16-19]). Badalian and Bakker in
their recent work [19] calculated and predicted the number as $\Delta _{{\rm %
hfs}}{\rm (2S,}${\rm theory}${\rm )=}57\pm 8$ ${\rm MeV}$ giving $M(\eta _{c}%
{\rm (2S})=$3630$\pm 8$ ${\rm MeV.}$ Clearly, the precision of the
experiments [2] requires a very substantial improvement to be sensitive to
the bound-state mass differences between the various calculations. In this
regard, from the global chi-square fitting values, there is a clear
preference for the Song-Lin $(\epsilon =1),$ Turin $(\epsilon =1/2)$ and
Cornell $(\epsilon =1/2)$ potentials, respectively. Therefore, we have found
that the Song-Lin $(\epsilon =1)$ potential is the best one fitting the $c%
\overline{c}$ and $b\overline{b}$ quarkonia.whereas the Cornell $(\epsilon
=1/2)$ potential is the worst one. Further, the Cornell $(\epsilon =1/2)$
potential seems to be the best fitting one for the $c\overline{c}$
quarkonium. In the\ pure scalar confinement $(\epsilon =1)$ couplings, we
have found that our fits are fairly good with level values and accurate to
several ${\rm MeV}.$ Further, the case of the vector confinement $(\epsilon
=0)$ interaction is being ruled out in our study since it gives the worst
fit to the spectra. We make the general remark as once $\beta $ value
increases, the $\epsilon $ value decreases. We have compared explicitly the
predicted and measured spin splitting energy for different $L$ states and
found that splitting approximation can be improved significantly by
increasing the quantum number $L.$

The deviations from experiment are more considerable. The calculation and
parameters are also model dependent [14]. Moreover, we tried another set of
parameters for the Cornell potential without permitting any additive
constant, that is, $V_{0}=0$ (cf. last column in Table I). We have also
found that the $\epsilon =1/2$ case is the best fitting one, in this work,
for the $c\overline{c}$ quarkonium and the worst one for the $b\overline{b}$
quarkonium (cf. Table V). It is clear that the coulomb-like parameter $A$ is
in accordance with the ideas of asymptotic freedom is expected for the
strong gauge-coupling constant of QCD [14]. For better fit to the quarkonium
spectra, the QCD coupling constant $\alpha _{s}(\mu ^{2})$ should be
dependent on the quark-flavour. The consideration of the variation of the
effective Coulomb interaction constant becomes especially essential for the $%
\Upsilon $ particle, for which $\alpha _{s}(\Upsilon )\neq \alpha _{s}(\psi
) $\footnote{%
For the best fit to the quarkonium spectra, the QCD coupling constant $%
\alpha _{s}(\mu ^{2})$ must be dependent on the quark flavour [11,19,35].
Motyka and Zalewiski [11] found $\frac{\alpha _{s}(m_{b}^{2})}{\alpha
_{s}(m_{c}^{2})}\simeq 11/18$ whereas Kiselev {\it et al.} [38] have found $%
\Delta M_{\Upsilon }(1S)=\frac{\alpha _{s}(\Upsilon )}{\alpha _{s}(\psi )}%
\Delta M_{\psi }(1S)$ with $\alpha _{s}(\Upsilon )/\alpha _{s}(\psi )\simeq
3/4.$}.

The calculated values of the pseudoscalar and vector decay constants of the $%
B_{c}$ meson using the nonrelativistic expression (56) are displayed in
Table VII. They are compared with the ones calculated using the
relativistic, nonrelativistic [1,35,38,39]. The radial wave function at the
origin has also been calculated in Table VIII and compared to the other
works available in literature [1,4]. These approximations have been
calculated without permitting any additive constant, that is, $V_{0}=0$ and
they also appear to be fairly good.\ \ \ \ \ \ \ \ \ \ \ \ \ \ \ \ \ \ \ \ \
\ \ \ \ \ \ \ \ \ \ \ \ \ \ \ \ \ \ \ \ \ \ \ \ \ \ \ \ \ \ \ \ \ \ \ \ \ \
\ \ \ \ \ \ \ \ \ \ \ \ \ \ \ \ \ \ \ \ \ \ \ \ \ \ \ \ \ \ \ \ \ \ \ \ \ \
\ \ \ \ \ \ \ \ \ \ \ \ \ \ \ \ \ \ \ \ \ \ \ \ \ \ \ \ \ \ \ \ \ \ \ \ \ \
\ \ \ \ \ \ \ \ \ \ 

\acknowledgments
S. M. Ikhdair gratefully acknowledges his wife, Oyoun, and also his son,
Musbah, for their love, encouragement and patience. Their encouragement
provided the necessary motivation to complete this work.\newpage

\appendix
\bigskip

\section{SLNET PARAMETERS FOR THE\ SCHR\"{O}DINGER EQUATION:}

Here, w{e list the analytic expressions of~ }$\alpha ${$^{(1)}$, }$\alpha ${$%
^{(2)}$, $\varepsilon _{i}$ and $\delta _{j}$ for the Schr\"{o}dinger
equation:}

\begin{eqnarray}
\alpha ^{(1)} &=&\frac{(1-a)(3-a)}{16\mu }+\left[ (1+2n_{r})~\bar{\varepsilon%
}_{2}~+3~(1+2n_{r}+2n_{r}^{2})~\bar{\varepsilon}_{4}\right] \;  \nonumber \\
&-&\omega ^{-1}\left[ ~\bar{\varepsilon}_{1}^{2}+6~(1+2n_{r})~\bar{%
\varepsilon}_{1}~\bar{\varepsilon}_{3}+~(11+30n_{r}+30n_{r}^{2})~\bar{%
\varepsilon}_{3}^{2}\right] ,  \label{A1}
\end{eqnarray}

\begin{eqnarray}
\alpha ^{(2)} &=&\left[ (1+2n_{r})~\bar{\delta}_{2}+3~(1+2n_{r}+2n_{r}^{2})~%
\bar{\delta}_{4}~+5~(3+8n_{r}+6n_{r}^{2}+4n_{r}^{3})~\bar{\delta}_{6}~\right.
\nonumber \\
&-&\omega ^{-1}~~(1+2n_{r})~\bar{\varepsilon}%
_{2}^{2}~+12~(1+2n_{r}+2n_{r}^{2})~\bar{\varepsilon}_{2}~\bar{\varepsilon}%
_{4}~+~2~\bar{\varepsilon}_{1}~\bar{\delta}_{1}  \nonumber \\
&+&2~(21+59n_{r}+51n_{r}^{2}+34n_{r}^{3})~\bar{\varepsilon}%
_{4}^{2}~+6~(1+2n_{r})~\bar{\varepsilon}_{1}~\bar{\delta}_{3}~  \nonumber \\
&+&30~(1+2n_{r}+2n_{r}^{2})~\bar{\varepsilon}_{1}~\bar{\delta}%
_{5}~+2~(11+30n_{r}+30n_{r}^{2})~\bar{\varepsilon}_{3}~\bar{\delta}_{3} 
\nonumber \\
&&+\left. \text{\ }10~(13\text{ }+\text{ }40n_{r}+42n_{r}^{2}\text{ }%
+28n_{r}^{3})~\bar{\varepsilon}_{3}~\bar{\delta}_{5}+6~(1+2n_{r})~\bar{%
\varepsilon}_{3}~\bar{\delta}_{1}\right]  \nonumber \\
\text{\ }~ &+&\omega ^{-2}\left[ ~4~\bar{\varepsilon}_{1}^{2}~\bar{%
\varepsilon}_{2}~+36~(1+2n_{r})~\bar{\varepsilon}_{1}~\bar{\varepsilon}_{2}~%
\bar{\varepsilon}_{3}+8~(11+30n_{r}+30n_{r}^{2})~\bar{\varepsilon}_{2}~\bar{%
\varepsilon}_{3}^{2}~\right.  \nonumber \\
&+&24~(1+2n_{r})~\bar{\varepsilon}_{1}^{2}~\bar{\varepsilon}%
_{4}+8~(31+78n_{r}+78n_{r}^{2})~\bar{\varepsilon}_{1}~\bar{\varepsilon}_{3}~%
\bar{\varepsilon}_{4}  \nonumber \\
&&+\left. 12~(57+189n_{r}+225n_{r}^{2}+150n_{r}^{3})~\bar{\varepsilon}%
_{3}^{2}~\bar{\varepsilon}_{4}\right]  \nonumber \\
&-&\omega ^{-3}\left[ ~8~\bar{\varepsilon}_{1}^{3}~\bar{\varepsilon}%
_{3}+108~(1+2n_{r})~\bar{\varepsilon}_{1}^{2}~\bar{\varepsilon}%
_{3}^{2}~+48~(11+30n_{r}+30n_{r}^{2})~\bar{\varepsilon}_{1}~\bar{\varepsilon}%
_{3}^{3}\right.  \nonumber \\
&+&\left. 30~(31+109n_{r}+141n_{r}^{2}+94n_{r}^{3})~\bar{\varepsilon}_{3}^{4}%
\right] ,~~~~~~~~~~~~  \label{A2}
\end{eqnarray}
where 
\begin{equation}
\bar{\varepsilon _{i}}~=~\frac{\varepsilon _{i}}{(4\mu \omega )^{i/2}}%
,~~~~i=1,2,3,4.~~~~~  \label{A3}
\end{equation}
and

\begin{equation}
\bar{\delta _{j}}=\frac{\delta _{j}}{(4\mu \omega )^{j/2}},~~~~j=1,2,3,4,5,6.
\label{A4}
\end{equation}

\begin{equation}
\varepsilon _{1}=\frac{(2-a)}{4\mu }~,{~~~}\varepsilon _{2}=-\frac{3}{8\mu }%
~(2-a),  \label{A5}
\end{equation}

\begin{equation}
\varepsilon _{3}=-\frac{1}{4\mu }~+\frac{r_{0}^{5}V^{\prime \prime \prime
}(r_{0})}{6Q};\text{ }~\varepsilon _{4}=\frac{5}{16\mu }+\frac{%
r_{0}^{6}~V^{\prime \prime \prime \prime }(r_{0})~}{24Q}  \label{A6}
\end{equation}

\begin{equation}
\delta _{1}=-\frac{(1-a)(3-a)}{8\mu };\text{ \ }\delta _{2}=\frac{%
3~(1-a)(3-a)}{16\mu },  \label{A7}
\end{equation}

\begin{equation}
\delta _{3}=\frac{(2-a)}{2\mu }~;{~~~}\delta _{4}=-\frac{5~(2-a)}{8\mu },
\label{A8}
\end{equation}

\begin{equation}
\delta _{5}=-\frac{3}{8\mu }+\frac{r_{0}^{7}~V^{\prime \prime \prime \prime
\prime }(r_{0})~}{120Q};\text{ \ }\delta _{6}=\frac{7}{16\mu }+\frac{%
r_{0}^{8}~V^{\prime \prime \prime \prime \prime \prime }(r_{0})~}{720Q}.
\label{A9}
\end{equation}

\section{THE SPIN-CORRECTION TERMS:}

For parastates $(S=0)$ case we have:

\begin{equation}
J=L
\end{equation}
For triplet $(S=1)$ case we have the following:

\begin{equation}
J=\left\{ 
\begin{array}{l}
L-1,\text{ }{\bf S}\cdot {\bf L}=-(L+1)~ \\ 
L,~{\bf S}\cdot {\bf L}=-1 \\ 
L+1,\text{ }{\bf S}\cdot {\bf L}=L
\end{array}
\right.
\end{equation}
The independent operators ${\bf S}_{1}\cdot {\bf S}_{2},$:$\left( {\bf S}%
_{1}\pm {\bf S}_{2}\right) \cdot {\bf L}$ and $T:$

\begin{equation}
\left\langle {\bf S}_{1}\cdot {\bf S}_{2}\right\rangle =\left\{ 
\begin{array}{l}
-3/4,\text{ for spin singlets }S=0,\text{ } \\ 
+1/4,\text{ for spin triplets }S=1.
\end{array}
\right.
\end{equation}

\begin{equation}
\left\langle {\bf S}\cdot {\bf L}\right\rangle =\left\{ 
\begin{array}{l}
0,\text{ for spin singlets }S=0,\text{ } \\ 
\frac{1}{2}\left[ J(J+1)-L(L+1)-2\right] ,\text{ for spin triplets }S=1.
\end{array}
\right.
\end{equation}

\begin{equation}
({\bf S}_{1}\cdot \widehat{{\bf r}}{\bf S}_{2}\cdot \widehat{{\bf r}}%
)u_{J}(r)Y_{J,0,J}(\widehat{{\bf r}})=-\frac{1}{4}u_{J}(r)Y_{J,0,J}(\widehat{%
{\bf r}}),
\end{equation}

\begin{equation}
({\bf S}_{1}\cdot {\bf S}_{2})Y_{J,S,L}^{m}(\widehat{{\bf r}})=\frac{1}{2}%
\left[ S(S+1)-S_{1}(S_{1}+1)-S_{2}(S_{2}+1)\right] Y_{J,S,L}^{m}(\widehat{%
{\bf r}}),
\end{equation}

\begin{equation}
\left( {\bf S}_{1}+{\bf S}_{2}\right) \cdot {\bf L}Y_{J,S,L}^{m}(\widehat{%
{\bf r}}){\bf =}\frac{1}{2}\left[ J(J+1)-L(L+1)-S(S+1)\right] Y_{J,S,L}^{m}(%
\widehat{{\bf r}}),
\end{equation}

\begin{equation}
\left( {\bf S}_{1}-{\bf S}_{2}\right) \cdot {\bf L}Y_{J,S,L}^{m}(\widehat{%
{\bf r}}){\bf =}\sqrt{\frac{1}{10}\left[ 2L+3)(2L-1)\right] }\delta
_{J,L}\left( \delta _{S,0}Y_{J,1,L}^{m}(\widehat{{\bf r}})+\delta
_{S,1}Y_{J,0,L}^{m}(\widehat{{\bf r}})\right) ,
\end{equation}

\[
TY_{J,1,L}^{m}(\widehat{{\bf r}}){\bf =}\frac{1}{4}\delta
_{J,L}Y_{J,1,L}^{m}(\widehat{{\bf r}})-\frac{1}{4(2L-1)}\delta
_{J,L-1}Y_{J,1,L}^{m}(\widehat{{\bf r}})+\frac{1}{4(2L+3)}\delta
_{J,L+1}Y_{J,1,L}^{m}(\widehat{{\bf r}}) 
\]

\begin{equation}
-\frac{\sqrt{(L+1)(L+2)}}{2(2L+3)}\delta _{J,L+1}Y_{J,1,L+2}^{m}(\widehat{%
{\bf r}})-\frac{\sqrt{L(L-1)}}{2(2L-1)}\delta _{J,L-1}Y_{J,1,L-2}^{m}(%
\widehat{{\bf r}}).
\end{equation}

\bigskip \baselineskip= 2\baselineskip

\widetext

\begin{table}[tbp]
\caption{Fitted parameters of the class of static central potentials. }
\label{table1}
\begin{tabular}{lllll}
Parameters & Cornell\tablenotemark[1] & Song-Lin\tablenotemark[1] & Turin%
\tablenote{These parameter fits [14] are used to produce masses in Tables
II--IV.} & Cornell\tablenote{These parameter fits [4] are used to produce
masses in Tables V-VI.} \\ 
\tableline$m_{c}~$ & 1.840$~GeV$ & 1.820$~GeV$ & 1.790$~GeV$ & 1.3205$~GeV$
\\ 
$m_{b}~$ & 5.232$~GeV$ & 5.199$~GeV$ & 5.171$~GeV$ & 4.7485$~GeV$ \\ 
$A$ & 0.520 & 0.923$~GeV^{1/2}$ & 0.620$~GeV^{1/4}$ & 0.472 \\ 
$\kappa ~$ & 0.1756$~GeV^{2}$ & 0.511$~GeV^{3/2}$ & 0.304$~GeV^{7/4}$ & 0.191%
$~GeV^{2}$ \\ 
$V_{0}~$ & -0.8578$~GeV$ & -0.798$~GeV$ & -0.823$~GeV$ & 0$~GeV$%
\end{tabular}
\end{table}
\bigskip


\newpage

\bigskip \widetext
\begin{table}[tbp]
\caption{Heavy$-$meson mass spectra (in $MeV$) for the Cornell potential.}
\label{table10}
\begin{tabular}{llllllllllll}
State & $\epsilon =1$ & $\epsilon =1/2$ & $\epsilon =0$ & $\epsilon =1$ & $%
\epsilon =1/2$ & $\epsilon =0$ & $\epsilon =1$ & $\epsilon =1/2$ & $\epsilon
=0$ & [1] & [4] \\ 
\tableline & $c\overline{c}$ &  &  & $b\overline{b}$ &  &  & $c\overline{b}$
&  &  &  &  \\ 
1$^{1}S_{0}$ & 3068.0 & 3047.0 & 3025.6 & 9424.5 & 9419.4 & 9414.2 & 6314.6
& 6305.3 & 6295.9 & 6264 & 6286 \\ 
2$^{1}S_{0}$ & 3658.5 & 3647.3 & 3635.9 & 10023.7 & 10021.4 & 10019.1 & 
6888.0 & 6883.3 & 6878.5 & 6856 & 6882 \\ 
3$^{1}S_{0}$ & 4075.7 & 4067.6 & 4059.5 & 10370.2 & 10368.6 & 10367.0 & 
7271.2 & 7267.8 & 7264.5 & 7244 &  \\ 
4$^{1}S_{0}$ & 4426.2 & 4419.6 & 4413.1 & 10642.4 & 10641.2 & 10640.0 & 
7587.2 & 7584.6 & 7581.9 &  &  \\ 
$1^{1}P_{1}$ & 3487.8 & 3477.4 & 3467.0 & 9918.1 & 9916.0 & 9913.9 & 6742.9
& 6738.5 & 6734.2 & 6730 & 6737 \\ 
2$^{1}P_{1}$ & 3921.8 & 3914.0 & 3906.2 & 10266.9 & 10265.4 & 10263.9 & 
7137.8 & 7134.7 & 7131.5 & 7135 &  \\ 
$1^{1}D_{2}$ & 3766.0 & 3758.6 & 3751.1 & 10162.7 & 10161.3 & 10159.9 & 
7003.1 & 7000.1 & 6997.0 & 7009 & 7028 \\ 
2$^{1}D_{2}$ & 4142.4 & 4136.2 & 4130.0 & 10448.0 & 10446.9 & 10445.7 & 
7340.2 & 7337.7 & 7335.2 &  &  \\ 
$1^{3}P_{1}$ & 3480.5 & 3464.4 & 3448.1 & 9910.1 & 9906.9 & 9903.7 & 6736.1
& 6726.8 & 6717.8 & 6736 & 6760 \\ 
2$^{3}P_{1}$ & 3920.6 & 3908.8 & 3896.9 & 10264.1 & 10261.9 & 10259.6 & 
7136.4 & 7129.6 & 7122.8 & 7142 &  \\ 
$1^{3}D_{2}$ & 3768.3 & 3757.1 & 3745.9 & 10161.8 & 10159.7 & 10157.5 & 
7004.2 & 6997.8 & 6991.4 & 7012 & 7028 \\ 
2$^{3}D_{2}$ & 4144.7 & 4135.4 & 4126.1 & 10447.5 & 10445.8 & 10444.1 & 
7341.4 & 7336.2 & 7330.9 &  &  \\ 
1$^{3}S_{1}$ & 3068.0 & 3074.9 & 3081.8 & 9424.5 & 9426.3 & 9428.0 & 6314.6
& 6317.7 & 6320.7 & 6337 & 6341 \\ 
2$^{3}S_{1}$ & 3658.5 & 3662.2 & 3665.9 & 10023.7 & 10024.5 & 10025.3 & 
6888.0 & 6889.5 & 6891.1 & 6899 & 6914 \\ 
3$^{3}S_{1}$ & 4075.7 & 4078.4 & 4081.1 & 10370.2 & 10370.7 & 10371.2 & 
7271.2 & 7272.3 & 7273.4 & 7280 &  \\ 
4$^{3}S_{1}$ & 4426.2 & 4428.3 & 4430.8 & 10642.4 & 10642.9 & 10643.3 & 
7587.2 & 7588.1 & 7589.0 &  &  \\ 
$1^{3}P_{2}$ & 3496.0 & 3522.7 & 3544.7 & 9928.7 & 9933.3 & 9937.9 & 6757.4
& 6765.8 & 6777.4 & 6747 & 6772 \\ 
2$^{3}P_{2}$ & 3926.3 & 3943.7 & 3962.8 & 10271.2 & 10274.6 & 10277.9 & 
7141.2 & 7150.3 & 7159.3 & 7153 &  \\ 
$1^{3}D_{3}$ & 3765.4 & 3796.4 & 3827.0 & 10166.3 & 10172.2 & 10178.1 & 
7003.1 & 7019.4 & 7035.7 & 7005 & 7032 \\ 
2$^{3}D_{3}$ & 4141.7 & 4168.8 & 4192.3 & 10450.3 & 10455.2 & 10460.0 & 
7339.4 & 7353.0 & 7366.5 &  &  \\ 
$1^{3}P_{0}$ & 3419.1 & 3365.0 & 3307.1 & 9874.7 & 9864.9 & 9855.0 & 6700.1
& 6673.6 & 6646.4 & 6700 & 6701 \\ 
2$^{3}P_{0}$ & 3899.8 & 3866.3 & 3832.1 & 10252.2 & 10246.0 & 10239.7 & 
7124.0 & 7106.5 & 7088.8 & 7108 &  \\ 
$1^{3}D_{1}$ & 3761.7 & 3716.4 & 3669.9 & 10154.9 & 10146.5 & 10138.1 & 
7000.3 & 6976.4 & 6952.1 & 7012 & 7019 \\ 
2$^{3}D_{1}$ & 4141.3 & 4104.3 & 4066.7 & 10443.1 & 10436.3 & 10429.6 & 
7339.6 & 7320.1 & 7300.4 &  & 
\end{tabular}
\end{table}

\bigskip \widetext
\begin{table}[tbp]
\caption{Heavy$-$meson mass spectra (in $MeV$) for the Song-Lin potential.}
\begin{tabular}{llllllllllll}
State & $\epsilon =1$ & $\epsilon =1/2$ & $\epsilon =0$ & $\epsilon =1$ & $%
\epsilon =1/2$ & $\epsilon =0$ & $\epsilon =1$ & $\epsilon =1/2$ & $\epsilon
=0$ & [35] & [38] \\ 
\tableline & $c\overline{c}$ &  &  & $b\overline{b}$ &  &  & $c\overline{b}$
&  &  &  &  \\ 
1$^{1}S_{0}$ & 3020.8 & 2991.4 & 2960.1 & 9417.7 & 9410.3 & 9402.7 & 6279.1
& 6266.4 & 6253.3 & 6270 & 6253 \\ 
2$^{1}S_{0}$ & 3634.3 & 3624.8 & 3615.0 & 10010.5 & 10008.0 & 10005.5 & 
6870.4 & 6866.1 & 6861.6 & 6835 & 6867 \\ 
3$^{1}S_{0}$ & 3983.6 & 3978.2 & 3972.7 & 10334.1 & 10332.7 & 10331.3 & 
7206.7 & 7204.2 & 7201.7 & 7193 &  \\ 
4$^{1}S_{0}$ & 4242.6 & 4238.9 & 4235.3 & 10567.8 & 10566.8 & 10565.9 & 
7454.3 & 7452.6 & 7451.0 &  &  \\ 
$1^{1}P_{1}$ & 3488.5 & 3480.6 & 3472.5 & 9879.3 & 9877.2 & 9875.1 & 6730.2
& 6726.6 & 6722.9 & 6734 & 6717 \\ 
2$^{1}P_{1}$ & 3873.9 & 3869.1 & 3864.2 & 10239.2 & 10237.9 & 10236.7 & 
7103.0 & 7100.7 & 7098.5 & 7126 & 7113 \\ 
$1^{1}D_{2}$ & 3761.6 & 3757.2 & 3752.9 & 10141.7 & 10140.6 & 10139.5 & 
6996.6 & 6994.6 & 6992.6 & 7077 & 7001 \\ 
2$^{1}D_{2}$ & 4061.1 & 4058.0 & 4054.8 & 10413.1 & 10412.3 & 10411.5 & 
7283.7 & 7282.3 & 7280.8 &  &  \\ 
$1^{3}P_{1}$ & 3483.3 & 3466.7 & 3449.6 & 9875.1 & 9870.8 & 9866.5 & 6725.7
& 6715.1 & 6704.2 & 6749 & 6729 \\ 
2$^{3}P_{1}$ & 3872.4 & 3862.5 & 3852.5 & 10237.4 & 10234.9 & 10232.4 & 
7101.6 & 7095.2 & 7088.8 & 7145 & 7124 \\ 
$1^{3}D_{2}$ & 3762.2 & 3753.4 & 3744.6 & 10141.0 & 10138.8 & 10136.5 & 
6996.8 & 6991.2 & 6985.6 & 7079 & 7016 \\ 
2$^{3}D_{2}$ & 4061.8 & 4055.5 & 4049.1 & 10412.7 & 10411.1 & 10409.5 & 
7284.0 & 7280.0 & 7276.0 &  &  \\ 
1$^{3}S_{1}$ & 3081.8 & 3089.3 & 3096.7 & 9447.8 & 9450.0 & 9452.2 & 6313.8
& 6317.5 & 6321.1 & 6332 & 6317 \\ 
2$^{3}S_{1}$ & 3646.0 & 3649.0 & 3652.0 & 10015.8 & 10016.7 & 10017.5 & 
6877.0 & 6878.5 & 6879.9 & 6881 & 6902 \\ 
3$^{3}S_{1}$ & 3988.1 & 3989.9 & 3991.6 & 10336.1 & 10336.6 & 10337.0 & 
7209.2 & 7210.0 & 7210.8 & 7235 &  \\ 
4$^{3}S_{1}$ & 4245.0 & 4246.2 & 4247.4 & 10568.8 & 10569.1 & 10569.4 & 
7455.6 & 7456.2 & 7456.7 &  &  \\ 
$1^{3}P_{2}$ & 3509.1 & 3533.0 & 3553.3 & 9892.1 & 9898.0 & 9898.5 & 6746.0
& 6758.0 & 6767.7 & 6762 & 6743 \\ 
2$^{3}P_{2}$ & 3885.5 & 3900.1 & 3912.1 & 10245.3 & 10248.9 & 10252.4 & 
7110.0 & 7118.0 & 7125.9 & 7156 & 7134 \\ 
$1^{3}D_{3}$ & 3771.0 & 3794.3 & 3816.3 & 10147.9 & 10153.9 & 10159.9 & 
7002.3 & 7016.1 & 7027.0 & 7081 & 7007 \\ 
2$^{3}D_{3}$ & 4067.6 & 4084.8 & 4100.6 & 10417.2 & 10421.4 & 10425.7 & 
7287.3 & 7297.3 & 7308.4 &  &  \\ 
$1^{3}P_{0}$ & 3422.7 & 3360.6 & 3288.7 & 9849.6 & 9836.2 & 9822.4 & 6693.1
& 6661.1 & 6627.0 & 6699 & 6683 \\ 
2$^{3}P_{0}$ & 3847.6 & 3816.4 & 3783.1 & 10226.1 & 10218.6 & 10211.1 & 
7087.6 & 7070.1 & 7052.0 & 7091 & 7088 \\ 
$1^{3}D_{1}$ & 3746.1 & 3708.2 & 3667.9 & 10132.5 & 10123.2 & 10113.8 & 
6988.0 & 6966.2 & 6943.5 & 7072 & 7008 \\ 
2$^{3}D_{1}$ & 4051.6 & 4025.0 & 3997.2 & 10407.3 & 10400.8 & 10394.2 & 
7278.6 & 7263.2 & 7247.3 &  & 
\end{tabular}
\end{table}

\bigskip \widetext
\begin{table}[tbp]
\caption{Heavy$-$meson mass spectra (in $MeV$) for the Turin potential.}
\begin{tabular}{llllllllllll}
State & $\epsilon =1$ & $\epsilon =1/2$ & $\epsilon =0$ & $\epsilon =1$ & $%
\epsilon =1/2$ & $\epsilon =0$ & $\epsilon =1$ & $\epsilon =1/2$ & $\epsilon
=0$ & [1] & [4] \\ 
\tableline & $c\overline{c}$ &  &  & $b\overline{b}$ &  &  & $c\overline{b}$
&  &  &  &  \\ 
1$^{1}S_{0}$ & 3041.9 & 3014.6 & 2986.1 & 9418.0 & 9411.3 & 9404.6 & 6290.5
& 6278.7 & 6266.7 & 6264 & 6286 \\ 
2$^{1}S_{0}$ & 3653.5 & 3642.0 & 3630.3 & 10005.1 & 10002.5 & 10000.0 & 
6877.3 & 6872.3 & 6867.3 & 6856 & 6882 \\ 
3$^{1}S_{0}$ & 4046.8 & 4039.3 & 4031.7 & 10343.3 & 10341.7 & 10340.1 & 
7244.9 & 7241.8 & 7238.6 & 7244 &  \\ 
4$^{1}S_{0}$ & 4360.3 & 4354.7 & 4349.1 & 10601.6 & 10600.4 & 10599.3 & 
7534.5 & 7532.2 & 7529.8 &  &  \\ 
$1^{1}P_{1}$ & 3489.4 & 3479.2 & 3468.8 & 9881.0 & 9878.7 & 9876.4 & 6729.2
& 6724.8 & 6720.4 & 6730 & 6737 \\ 
2$^{1}P_{1}$ & 3910.7 & 3903.7 & 3896.6 & 10241.2 & 10239.7 & 10238.2 & 
7122.6 & 7119.6 & 7116.7 & 7135 &  \\ 
$1^{1}D_{2}$ & 3771.9 & 3765.4 & 3758.9 & 10136.7 & 10135.3 & 10134.0 & 
6997.7 & 6995.0 & 6992.2 & 7009 & 7028 \\ 
2$^{1}D_{2}$ & 4120.7 & 4115.6 & 4110.5 & 10421.9 & 10420.8 & 10419.8 & 
7319.2 & 7317.1 & 7315.0 &  &  \\ 
$1^{3}P_{1}$ & 3484.5 & 3466.2 & 3447.6 & 9875.5 & 9871.5 & 9867.5 & 6724.6
& 6713.6 & 6702.4 & 6736 & 6760 \\ 
2$^{3}P_{1}$ & 3910.1 & 3898.0 & 3885.6 & 10239.1 & 10236.6 & 10234.0 & 
7121.7 & 7114.4 & 7107.1 & 7142 &  \\ 
$1^{3}D_{2}$ & 3774.1 & 3762.9 & 3751.6 & 10136.1 & 10133.7 & 10131.3 & 
6999.0 & 6992.3 & 6985.5 & 7012 & 7028 \\ 
2$^{3}D_{2}$ & 4122.7 & 4114.0 & 4105.2 & 10421.5 & 10419.7 & 10417.9 & 
7320.4 & 7315.3 & 7310.0 &  &  \\ 
1$^{3}S_{1}$ & 3075.4 & 3083.3 & 3091.1 & 9441.3 & 9443.3 & 9445.4 & 6311.8
& 6315.4 & 6319.0 & 6337 & 6341 \\ 
2$^{3}S_{1}$ & 3659.0 & 3662.8 & 3666.5 & 10008.0 & 10008.8 & 10009.7 & 
6880.5 & 6882.1 & 6883.8 & 6899 & 6914 \\ 
3$^{3}S_{1}$ & 4048.9 & 4051.3 & 4053.8 & 10344.3 & 10344.8 & 10345.3 & 
7246.1 & 7247.2 & 7248.2 & 7280 &  \\ 
4$^{3}S_{1}$ & 4361.4 & 4363.3 & 4365.1 & 10602.0 & 10602.4 & 10602.8 & 
7535.1 & 7535.9 & 7536.7 &  &  \\ 
$1^{3}P_{2}$ & 3502.7 & 3529.7 & 3553.5 & 9891.9 & 9897.5 & 9903.1 & 6740.0
& 6753.8 & 6767.3 & 6747 & 6772 \\ 
2$^{3}P_{2}$ & 3917.7 & 3936.8 & 3952.7 & 10246.2 & 10249.9 & 10253.7 & 
7127.1 & 7136.6 & 7146.0 & 7153 &  \\ 
$1^{3}D_{3}$ & 3773.9 & 3804.8 & 3834.2 & 10141.1 & 10147.6 & 10154.1 & 
6999.2 & 7016.0 & 7032.0 & 7005 & 7032 \\ 
2$^{3}D_{3}$ & 4122.8 & 4145.7 & 4168.4 & 10424.7 & 10429.7 & 10434.7 & 
7319.7 & 7332.8 & 7345.8 &  &  \\ 
$1^{3}P_{0}$ & 3426.8 & 3362.9 & 3291.7 & 9847.9 & 9835.7 & 9823.3 & 6692.8
& 6661.0 & 6627.6 & 6700 & 6701 \\ 
2$^{3}P_{0}$ & 3888.3 & 3852.1 & 3814.3 & 10227.9 & 10220.5 & 10213.1 & 
7109.3 & 7089.9 & 7070.1 & 7108 &  \\ 
$1^{3}D_{1}$ & 3764.8 & 3718.1 & 3669.3 & 10129.0 & 10119.4 & 10109.6 & 
6994.1 & 6968.6 & 6942.4 & 7012 & 7019 \\ 
2$^{3}D_{1}$ & 4117.3 & 4081.6 & 4044.9 & 10416.9 & 10409.7 & 10402.4 & 
7317.7 & 7298.3 & 7278.4 &  & 
\end{tabular}
\end{table}

\bigskip

\bigskip \widetext
\begin{table}[tbp]
\caption{$c\overline{c}$ and $b\overline{b}$ mass spectra (in $MeV$) using
the Cornell potential. }
\label{table11}
\begin{tabular}{llllllllllll}
State & Meson\tablenote{Same parameter fits of Ref. [4] in Table I with
$V_{0}=0$.} & [35] & $\epsilon =1$ & $\epsilon =1/2$ & $\epsilon =0$ & Meson
& [35] & $\epsilon =1$ & $\epsilon =1/2$ & $\epsilon =0$ &  \\ 
\tableline1$^{1}S_{0}$ & $\eta _{c}$ & 2979 & 3068.5 & 3031.4 & 2993.3 & $%
\eta _{b}$ & 9400 & 9447.4 & 9441.3 & 9435.3 &  \\ 
2$^{1}S_{0}$ & $\eta _{c}^{\prime }$ & 3588 & 3704.7 & 3683.8 & 3662.6 & $%
\eta _{b}^{\prime }$ & 9993 & 10021.5 & 10018.6 & 10015.7 &  \\ 
3$^{1}S_{0}$ & $\eta _{c}^{\prime \prime }$ & 3991 & 4177.2 & 4161.8 & 4146.3
& $\eta _{b}^{\prime \prime }$ & 10328 & 10378.4 & 10376.4 & 10374.4 &  \\ 
4$^{1}S_{0}$ &  &  & 4580.6 & 4568.1 & 4555.6 &  &  & 10665.8 & 10664.2 & 
10662.6 &  \\ 
$1^{1}P_{1}$ & $h_{c}$ & 3526 & 3497.6 & 3478.1 & 3458.4 & $h_{b}$ & 9901 & 
9899.7 & 9897.1 & 9894.5 &  \\ 
2$^{1}P_{1}$ & $h_{c}^{\prime }$ & 3945 & 3993.2 & 3978.4 & 3963.5 & $%
h_{b}^{\prime }$ & 10261 & 10263.3 & 10261.4 & 10259.5 &  \\ 
$1^{1}D_{2}$ &  & 3811 & 3806.7 & 3792.6 & 3778.4 &  & 10158 & 10147.2 & 
10145.4 & 10143.6 &  \\ 
2$^{1}D_{2}$ &  &  & 4242.9 & 4231.0 & 4219.2 &  &  & 10451.0 & 10449.5 & 
10448.1 &  \\ 
$1^{3}P_{1}$ & $\chi _{c1}$ & 3510 & 3496.0 & 3465.7 & 3434.9 & $\chi _{b1}$
& 9892 & 9893.2 & 9889.2 & 9885.2 &  \\ 
2$^{3}P_{1}$ & $\chi _{c1}^{\prime }$ & 3929 & 3996.7 & 3974.2 & 3951.4 & $%
\chi _{b1}^{\prime }$ & 10255 & 10261.0 & 10258.2 & 10255.3 &  \\ 
$1^{3}D_{2}$ &  & 3813 & 3814.3 & 3793.0 & 3771.5 &  & 10158 & 10146.7 & 
10144.0 & 10141.2 &  \\ 
2$^{3}D_{2}$ &  &  & 4249.6 & 4231.9 & 4214.0 &  &  & 10450.8 & 10448.6 & 
10446.4 &  \\ 
1$^{3}S_{1}$ & $J/\psi $ & 3096 & 3068.5 & 3080.6 & 3092.7 & $\Upsilon $ & 
9460 & 9447.4 & 9449.4 & 9451.4 &  \\ 
2$^{3}S_{1}$ & $\psi ^{\prime }$ & 3686 & 3704.7 & 3711.6 & 3718.5 & $%
\Upsilon ^{\prime }$ & 10023 & 10021.5 & 10022.4 & 10023.4 &  \\ 
3$^{3}S_{1}$ & $\psi ^{\prime \prime }$ & 4088 & 4177.2 & 4182.3 & 4187.4 & $%
\Upsilon ^{\prime \prime }$ & 10355 & 10378.4 & 10379.0 & 10379.7 &  \\ 
4$^{3}S_{1}$ & $\psi ^{\prime \prime \prime }$ &  & 4580.6 & 4584.8 & 4588.9
& $\Upsilon ^{\prime \prime \prime }$ &  & 10665.8 & 10666.3 & 10666.8 &  \\ 
$1^{3}P_{2}$ & $\chi _{c2}$ & 3556 & 3505.2 & 3547.6 & 3589.1 & $\chi _{b2}$
& 9913 & 9908.9 & 9914.7 & 9920.5 &  \\ 
2$^{3}P_{2}$ & $\chi _{c2}^{\prime }$ & 3972 & 3994.0 & 4027.1 & 4059.8 & $%
\chi _{b2}^{\prime }$ & 10268 & 10267.1 & 10271.4 & 10275.7 &  \\ 
$1^{3}D_{3}$ &  & 3815 & 3796.9 & 3855.8 & 3913.4 &  & 10162 & 10150.0 & 
10157.6 & 10165.2 &  \\ 
2$^{3}D_{3}$ &  &  & 4233.3 & 4283.1 & 4331.9 &  &  & 10452.8 & 10459.0 & 
10465.2 &  \\ 
$1^{3}P_{0}$ & $\chi _{c0}$ & 3424 & 3430.9 & 3327.5 & 3209.5 & $\chi _{b0}$
& 9863 & 9862.9 & 9851.0 & 9839 &  \\ 
2$^{3}P_{0}$ & $\chi _{c0}^{\prime }$ & 3854 & 3975.7 & 3911.7 & 3845.3 & $%
\chi _{b0}^{\prime }$ & 10234 & 10249.9 & 10242.0 & 10234.1 &  \\ 
$1^{3}D_{1}$ &  & 3798 & 3815.4 & 3728.7 & 3638.2 &  & 10153 & 10140.7 & 
10130.0 & 10119.2 &  \\ 
2$^{3}D_{1}$ &  &  & 4252.9 & 4181.8 & 4108.7 &  &  & 10447.0 & 10438.3 & 
10429.5 & 
\end{tabular}
\end{table}

\widetext
\begin{table}[tbp]
\caption{$B_{c}$ meson mass spectrum (in $MeV$) for the Cornell potential.}
\label{table14}
\begin{tabular}{lllllllll}
State\tablenote{Same parameter fits of Ref. [4] in Table I with $V_{0}=0$.}
& $\epsilon =1$ & $\epsilon =1/2$ & $\epsilon =0$ & [35] & [38] & [1] & [4]
& [40] \\ 
\tableline1$^{1}S_{0}$ & 6338.7 & 6325.7 & 6312.6 & 6270 & 6253 & 6264 & 6286
& $\geq 6219.6$ \\ 
2$^{1}S_{0}$ & 6930.5 & 6923.5 & 6916.5 & 6835 & 6867 & 6856 & 6882 &  \\ 
3$^{1}S_{0}$ & 7352.2 & 7347.1 & 7342.0 & 7193 &  & 7244 &  &  \\ 
4$^{1}S_{0}$ & 7707.2 & 7703.2 & 7699.1 &  &  &  &  &  \\ 
$1^{1}P_{1}$ & 6756.0 & 6749.5 & 6743.0 & 6734 & 6717 & 6730 & 6737 & $\geq
6701.2$ \\ 
2$^{1}P_{1}$ & 7195.2 & 7190.3 & 7185.5 & 7126 & 7113 & 7135 &  &  \\ 
$1^{1}D_{2}$ & 7036.3 & 7031.7 & 7027.1 & 7077 & 7001 & 7009 & 7028 &  \\ 
2$^{1}D_{2}$ & 7418.1 & 7414.3 & 7410.4 &  &  &  &  &  \\ 
$1^{3}P_{1}$ & 6753.6 & 6737.4 & 6720.9 & 6749 & 6729 & 6736 & 6760 & $\geq
6701.2$ \\ 
2$^{3}P_{1}$ & 7196.9 & 7184.9 & 7172.8 & 7145 & 7124 & 7142 &  &  \\ 
$1^{3}D_{2}$ & 7040.9 & 7029.5 & 7018.1 & 7079 & 7016 & 7012 & 7028 &  \\ 
2$^{3}D_{2}$ & 7422.2 & 7412.8 & 7403.4 &  &  &  &  &  \\ 
1$^{3}S_{1}$ & 6338.7 & 6342.9 & 6347.2 & 6332 & 6317 & 6337 & 6341 & $\geq
6278.6$ \\ 
2$^{3}S_{1}$ & 6930.5 & 6932.8 & 6935.1 & 6881 & 6902 & 6899 & 6914 &  \\ 
3$^{3}S_{1}$ & 7352.2 & 7353.8 & 7355.5 & 7235 &  & 7280 &  &  \\ 
4$^{3}S_{1}$ & 7707.2 & 7708.6 & 7710.0 &  &  &  &  &  \\ 
$1^{3}P_{2}$ & 6761.2 & 6780.6 & 6799.8 & 6762 & 6743 & 6747 & 6772 & $\geq
6734.7$ \\ 
2$^{3}P_{2}$ & 7195.5 & 7211.0 & 7226.3 & 7156 & 7134 & 7153 &  &  \\ 
$1^{3}D_{3}$ & 7029.7 & 7057.3 & 7085.5 & 7081 & 7007 & 7005 & 7032 &  \\ 
2$^{3}D_{3}$ & 7411.8 & 7435.6 & 7458.6 &  &  &  &  &  \\ 
$1^{3}P_{0}$ & 6724.1 & 6680.2 & 6634.5 & 6699 & 6683 & 6700 & 6701 & $\geq
6638.6$ \\ 
2$^{3}P_{0}$ & 7187.5 & 7157.5 & 7127.0 & 7091 & 7088 & 7108 &  &  \\ 
$1^{3}D_{1}$ & 7043.6 & 7002.3 & 6960.1 & 7072 & 7008 & 7012 & 7019 &  \\ 
2$^{3}D_{1}$ & 7425.9 & 7391.9 & 7357.4 &  &  &  &  & 
\end{tabular}
\end{table}

\bigskip

\bigskip

\mediumtext

\bigskip \widetext

\bigskip

\begin{table}[tbp]
\caption{Pseudoscalar and vector decay constants $%
(f_{P}=f_{B_{c}},~f_{V}=f_{B_{c}^{\ast }})$ of the $B_{c}$ meson (in $MeV$)
using the Cornell potential.}
\label{table7}
\begin{tabular}{llllllllll}
Constants\tablenote{For parameter fits we cite Ref. [4].} & SLNET%
\tablenote{Here ($\epsilon=0$).} & SLNET\tablenote{Here ($\epsilon=1/2$).} & 
SLNET\tablenote{Here ($\epsilon=1$).} & Rel[35] & [35] & [1] & [38] & [4] & 
[39] \\ 
\tableline$f_{B_{c}}$ & 511.4 & 503.2 & 495.2 & 433 & 562 & 479-687 & 460$%
\pm 60$ & 517 & 420$\pm 13$ \\ 
$f_{B_{c}^{\ast }}$ & 490.0 & 492.6 & 495.2 & 503 & 562 & 479-687 & 460$\pm
60$ & 517 & -
\end{tabular}
\end{table}

\bigskip \mediumtext

\bigskip 
\begin{table}[tbp]
\caption{The radial wave function at the origin (in $GeV^{3})$ calculated in
our model and by the other authors using the Cornell potential.}
\label{table8}
\begin{tabular}{llllllll}
Level\tablenote{For parameter fits we cite Ref. [4].} & SLNET\tablenote{Here
($\epsilon=0$).} & SLNET\tablenote{Here ($\epsilon=1/2$).} & SLNET%
\tablenote{Here ($\epsilon=1$).} & Martin & [1] & [4] & [1]\tablenote{For
the 1S level.} \\ 
\tableline$\left| R_{B_{c}}(0)\right| ^{2}$ & 1.729 & 1.677 & 1.628 & 1.716
& 1.638 & 1.81 & 1.508-3.102 \\ 
$\left| R_{B_{c}^{\ast }}(0)\right| ^{2}$ & 1.596 & 1.612 & 1.628 & - & - & -
& -
\end{tabular}
\end{table}

\bigskip

\bigskip \bigskip

\end{document}